\documentstyle[12pt]{article}
\def\be{\begin{equation}}
\def\ee{\end{equation}}
\def\bc{\begin{center}}
\def\ec{\end{center}}
\def\bea{\begin{eqnarray}}
\def\eea{\end{eqnarray}}
\def\atu{{\alpha_t^U}}
\def\cm{{\cal M}}
\def\dvc{{\Delta V_{cosm}}}
\def\ov{\overline}
\def\gev{{\rm \; GeV}}
\def\tev{{\rm \; TeV}}
\def\mpl{{M_{\rm P}}}
\def\msu{{M_{\rm SUSY}}}
\def\simlt{\stackrel{<}{{}_\sim}}

\def\cg{{\cal G}}
\def\str{{\rm \; Str \;}}

\def\zzbar{{(z, \overline{z})}}

\catcode`@=11
\def\marginnote#1{}
\newcount\hour
\newcount\minute
\newtoks\amorpm
\hour=\time\divide\hour by60
\minute=\time{\multiply\hour by60 \global\advance\minute by-\hour}
\edef\standardtime{{\ifnum\hour<12 \global\amorpm={am}%
        \else\global\amorpm={pm}\advance\hour by-12 \fi
        \ifnum\hour=0 \hour=12 \fi
        \number\hour:\ifnum\minute<10 0\fi\number\minute\the\amorpm}}
\edef\militarytime{\number\hour:\ifnum\minute<10 0\fi\number\minute}
\def\draftlabel#1{{\@bsphack\if@filesw {\let\thepage\relax
   \xdef\@gtempa{\write\@auxout{\string
      \newlabel{#1}{{\@currentlabel}{\thepage}}}}}\@gtempa
   \if@nobreak \ifvmode\nobreak\fi\fi\fi\@esphack}
        \gdef\@eqnlabel{#1}}
\def\@eqnlabel{}
\def\@vacuum{}
\def\draftmarginnote#1{\marginpar{\raggedright\scriptsize\tt#1}}
\def\draft{\oddsidemargin 0.0truein
        \def\@oddfoot{\sl preliminary draft \hfil
        \rm\thepage\hfil\sl\today\quad\militarytime}
        \let\@evenfoot\@oddfoot \overfullrule 3pt
        \let\label=\draftlabel
        \let\marginnote=\draftmarginnote
   \def\@eqnnum{(\theequation)\rlap{\kern\marginparsep\tt\@eqnlabel}%
\global\let\@eqnlabel\@vacuum}  }
\catcode`@=12
\begin{document}
\begin{titlepage}
\vspace*{-1cm}
\phantom{bla}
\hfill{CERN-TH.7325/94}
\\
\phantom{bla}
\hfill{hep-ph/9407215}
\vskip 2.0cm
\begin{center}
{\Large\bf Beyond the MSSM}
\end{center}
\vskip 1.5cm
\begin{center}
{\large Fabio Zwirner}\footnote{On leave from INFN, Sezione di
Padova,
Padua, Italy. Supported in part by the European Union under contract
No.~CHRX-CT92-0004.}
\\
\vskip .5cm
Theory Division, CERN \\
CH-1211 Geneva 23, Switzerland
\end{center}
\vskip 1cm
\begin{abstract}
\noindent
To increase the predictivity of the Minimal Supersymmetric Standard
Model (MSSM), one needs to go to an underlying, more fundamental
theory,
where at least some of the many MSSM parameters can be determined by
symmetries or by dynamics. Progress may come from
four-dimensional superstring solutions and their effective
supergravities. Summarizing some recent work [\ref{fkz}--\ref{kprz}],
we introduce a class of `large-hierarchy-compatible' (LHC) models
that could naturally embed a stable hierarchy $m_Z \simlt m_{3/2}
\ll \mpl$. We discuss how in LHC models one may determine: 1) the
explicit mass terms of the MSSM, as functions of the gravitino mass;
2) the scales of gauge and supersymmetry breaking, $m_Z$ and
$m_{3/2}$;
3) the heavy-fermion masses.
\end{abstract}
\vskip 1.5cm
\begin{center}
{\it
Based on talks given at: the Second IFT Workshop on Yukawa Couplings,
Gainesville, Florida, USA, 11--13 February 1994; the XXIXth
Rencontres
de Moriond, M\'eribel, France, 12--19 March 1994; the First
International
Conference on Phenomenology of Unification from Present to Future,
Rome,
Italy, 23--26 March 1994.
}
\end{center}
\vfill{
CERN-TH.7325/94
\newline
\noindent
June 1994}
\end{titlepage}
\setcounter{footnote}{0}
\vskip2truecm
\begin{center}
{\bf 1. Introduction}
\end{center}

The gauge hierarchy problem of the Standard Model (SM) is related to
the existence of quadratically divergent one-loop corrections to the
effective potential, proportional to
\be
\str \cm^2 (\phi) \equiv \sum_i (-1)^{2J_i} (2 J_i + 1) m_i^2 (\phi)
\, ,
\ee
where $\phi$ is the classical Higgs field and the index $i$ runs over
the states of the model, with field-dependent squared masses $m_i^2
(\phi)$ and spins $J_i$. Correspondingly, there are also
quadratically
divergent contributions to the Higgs mass, proportional to
$[\partial^2
\str \cm^2 / \partial \phi^2]$. Since in the SM both quantities
are generically non-vanishing, the natural scale of the SM Higgs mass
and of the corresponding VEV [if $SU(2) \times U(1)$ is broken] is
the
ultraviolet cut-off scale, e.g. the Planck scale: a ratio $m_Z /
\mpl \sim 10^{-16}$ is unstable versus perturbative quantum
corrections.

A partial solution of the gauge hierarchy problem is provided
by the Minimal Supersymmetric Standard Model (MSSM). In this
model [\ref{mssm}], supersymmetry breaking is parametrized by a
collection of explicit but soft mass parameters, such that,
denoting spin-0 fields with the generic symbol $z$,
$\str \cm^2 (z) = constant$,
and there are no field-dependent quadratic divergences. The scale
$\msu$ of the explicit MSSM mass terms acts as an effective cut-off
scale for the SM, and a sufficient condition to solve the gauge
hierarchy problem is to have $\msu \simlt 1 \tev$. Usually, the soft
mass terms are further constrained by assuming, as boundary
conditions
at some grand-unification scale $M_{\rm U} \sim 10^{16} \gev$, a
universal scalar mass $m_0$, a universal gaugino mass $m_{1/2}$, and
a universal cubic scalar coupling $A$. It should be kept in mind,
however,
that such assumptions are not based on fundamental symmetry
principles,
and that different sets of assumptions can also give
phenomenologically
viable models.

In addition to the stabilization of the hierarchy, another
attractive feature of the MSSM is the possibility of describing
the spontaneous breaking of the electroweak gauge symmetry as an
effect of radiative corrections. Thanks to the quantum
corrections associated with the large top-quark Yukawa coupling, the
effective potential of the MSSM gives a phenomenologically acceptable
vacuum, with $SU(2)_L \times U(1)_Y$ broken down to $U(1)_{em}$ and
$m_Z$
naturally of order $\msu$. For appropriate numerical assignments of
the
boundary conditions, one obtains a mass spectrum compatible with
present
experimental data.

Besides these virtues, the MSSM has a very unsatisfactory feature:
the numerical values of its explicit mass parameters must
be arbitrarily chosen `by hand' (the fact that also the gauge and
Yukawa couplings must be chosen `by hand' is as unsatisfactory as in
the Standard Model). This means a certain lack of predictivity, and
in particular does not provide any dynamical explanation for the
origin of the hierarchy $\msu \ll \mpl$, which is just assumed to be
there. To go further, one must have a model for spontaneous
supersymmetry breaking in the fundamental theory underlying the MSSM.
The only possible candidate for such a theory is $N=1$ supergravity,
where, in contrast with the case of global supersymmetry, the
spontaneous breaking of local supersymmetry is not incompatible with
vanishing vacuum energy. For spontaneous breaking on a flat
background,
the order parameter is the gravitino mass, $m_{3/2}$, and all the
explicit mass parameters of the MSSM are calculable (but
model-dependent)
functions of $m_{3/2}$.

When discussing the spontaneous breaking of $N=1$ supergravity,
one is faced with some hierarchy problems that are as serious as
the gauge hierarchy problem of the SM. In $N=1$ supergravity, the
gravitino mass $m_{3/2}$ depends
on the VEVs of the scalar fields, some of which typically have
masses of order $m_{3/2}$ or smaller, and there are in general
field-dependent quadratically divergent contributions to the
effective potential. This implies that a small ratio $m_{3/2}/
\mpl$ is generically unstable versus perturbative quantum
corrections. In addition, consistency with the flat background,
explicitly assumed in the standard formalism of Poincar\'e
supergravity, asks for the absence of ${\cal O}(m_{3/2}^2 \mpl^2)$
contributions to the vacuum energy when discussing physics at
scales $Q \sim m_Z \simlt m_{3/2}$. Already at the classical
level, this is a highly non-trivial requirement, since the scalar
potential of $N=1$ supergravity is not positive-semidefinite.
This problem is partially solved by the so-called `no-scale' models
[\ref{nscl},\ref{nsqu}], which naturally fit in the effective
supergravity theories derived from classical four-dimensional
vacua of the heterotic superstring. In these models, the classical
potential is manifestly positive-semidefinite, and all its minima
correspond to broken supersymmetry and vanishing vacuum energy, with
the gravitino mass sliding along an approximately flat
direction [\ref{nscl}]. The fact that the supersymmetry breaking
scale is classically undetermined provides the possibility of fixing
it via quantum corrections. If the latter, as computed in the
fundamental quantum theory of gravity, do not introduce ${\cal O}
(m_{3/2}^2 \mpl^2)$ contributions to the effective potential, then
one may obtain an exponentially suppressed gravitino mass, $m_{3/2}
\sim \exp [ - {\cal O}(1) / \alpha ] \mpl$, as a result of the
logarithmic quantum corrections in the low-energy effective
supergravity theory [\ref{nsqu}].

\begin{center}
{\bf 2. LHC supergravity models}
\end{center}

At the level of the effective $N=1$ supergravity, the viability of
the above program (and of other scenarios for the generation of the
hierarchy $m_{3/2} \ll \mpl$) is plagued by the possible existence
of quadratically divergent contributions to the vacuum energy,
proportional to
\be
\label{genstr}
\str \cm^2 \zzbar = 2 \, Q \zzbar \, m_{3/2}^2 \zzbar
\, ,
\ee
where [\ref{grk}], using here and in the following the
supergravity convention $\mpl=1$, and assuming for simplicity
$F$-breaking along a gauge singlet direction,
\be
\label{qexpr}
Q \zzbar  =  N_{TOT} - 1 - \cg^I  \zzbar
\left[ R_{I \ov{J}} \zzbar + F_{I \ov{J}} \zzbar \right]
\cg^{\ov{J}} \zzbar \, ,
\ee
\be
\label{ricci}
R_{I \ov{J}} \zzbar \equiv \partial_{I} \partial_{\ov{J}}
\log \det \cg_{M \ov{N}} \zzbar \, ,
\;\;\;\;\;
F_{I \ov{J}} \zzbar \equiv \partial_{I} \partial_{\ov{J}}
\log \det [{\rm Re \,} f_{ab} (z) ]^{-1} \, .
\ee
To interpret the previous formulae, we recall that the $N=1$
supergravity
Lagrangian is determined by two arbitrary functions: the K\"ahler
function
$\cg (z,\ov{z}) = K (z,\ov{z}) + \log |w(z)|^2$, where $K$ is the
K\"ahler
potential, whose second derivatives determine the kinetic terms for
the
fields in the chiral supermultiplets, and $w$ is the superpotential;
the
gauge kinetic function $f_{ab} (z)$, which determines the kinetic
terms
for the fields in the vector supermultiplets, and in particular the
gauge
coupling constants $g_{ab}^{-2} =  {\rm Re \,} f_{ab}$. In
eqs.~(\ref{qexpr}) and (\ref{ricci}), derivatives of the K\"ahler
function
are denoted by $\partial \cg / \partial z^I \equiv \partial_I \cg
\equiv \cg_I$ and $\partial \cg / \partial \ov{z}^{\ov{I}} \equiv
\partial_{\ov{I}} \cg \equiv \cg_{\ov{I}}$, the K\"ahler metric
is $\cg_{I \ov{J}} = \cg_{\ov{J} I}$, and the inverse K\"ahler
metric $\cg^{I \ov{J}}$ is used to define $\cg^I \equiv \cg^{I
\ov{J}}
\cg_{\ov{J}}$ and $\cg^{\ov{I}} \equiv \cg_{J}\cg^{J \ov{I}}$.
It is important to observe that both $R_{I \ov{J}}$ and $F_{I
\ov{J}}$
do not depend on the superpotential, but only depend on the metrics
for the chiral and gauge superfields. This allows for the possibility
that, for special geometrical properties of these two metrics, the
dimensionless quantity $Q \zzbar$ may turn out to be
field-independent
and hopefully vanishing.

In order to appreciate the geometrical meaning of the vanishing
of $Q \zzbar$, we present here a simple working example [\ref{fkz}].
Consider a model containing $N_{TOT} \equiv N_c + 3$ chiral
superfields, three gauge singlets $(T,U,S)$ and $N_c$ charged
fields $C_i$ ($i=1,\ldots,N_c$), with a gauge kinetic function
$f_{ab} = \delta_{ab} S$, a K\"ahler function
\be
\cg =
- 3 \log (T + \ov{T} - C_i \ov{C}_i)
- k \log (U + \ov{U}) - \log (S + \ov{S})
+ \log | w(C,U,S) |^2 \, ,
\ee
and a superpotential $w(C,U,S)$, which depends non-trivially on all
fields apart from the singlet field $T$. One can easily prove that,
thanks to the field identity $\cg^T \cg_T \equiv K^T K_T \equiv 3$,
the scalar potential of such a model is automatically positive
semidefinite, with a flat direction along the $T$-field. As long as
there are field configurations for which $w \ne 0$ with $\cg_S =
\cg_U = \cg_C = 0$, there are minima that preserve the gauge symmetry
but break supersymmetry with vanishing vacuum energy and $\cg_T \ne
0$.
The gauge coupling constant at the minimum is fixed
to the value $g^2 = ({\rm Re \, } S)^{-1}$, and the
VEV of the $U$ field is also fixed by the minimization condition,
whereas the gravitino mass $m_{3/2}^2= |w|^2 / [(S + \ov{S})
(T + \ov{T})^3 (U + \ov{U})^k]$ is classically undetermined, sliding
along the $T$ flat direction.  To compute $Q \zzbar$ in this model,
it is sufficient to realize that the Ricci tensors for the three
factor manifolds have the simple expressions ($I=0,1,\ldots,N_c$):
\be
R_{I \ov{J}} = {N_c+2 \over 3} \cg_{I \ov{J}} \, ,
\;\;\;\;\;
R_{S \ov{S}} = 2 \cg_{S \ov{S}} \, ,
 \;\;\;\;\;
R_{U \ov{U}} = {2 \over k} \cg_{U \ov{U}} \, ,
\ee
from which one finds, by just applying eqs.~(\ref{qexpr}) and
(\ref{ricci}), that $Q \zzbar \equiv 0$ at all minima of the
potential along the flat direction $T$, independently of the
details of the superpotential.

The previous example can be generalized [\ref{fkz}] to the case of
a supergravity model containing $N_{TOT}$ fields $z^I \equiv
( z^{\alpha}, z^i)$ and described, for small field fluctuations
around $\langle z^{i} \rangle = 0$, by the K\"ahler function
\be
\label{totkf}
\cg = - \log Y (r^{\alpha}) + \sum_A K^A_{i_A \ov{j}_A}
(r^{\alpha}) z^{i_A} \ov{z}^{\ov{j}_A}+ {1 \over 2} \sum_{A,B}
\left[ P_{i_A j_B}  (r^{\alpha}) z^{i_A} z^{j_B} + {\rm h.c.}
\right] + \log |w(z^{i})|^2  \, ,
\ee
depending on the fields $z^{\alpha}$ only via the real combinations
$r^\alpha \equiv z^\alpha + \ov{z}^{\ov{\alpha}}$. The K\"ahler
potential and the gauge kinetic function are assumed  to have
the scaling properties
\be
r^\alpha Y_\alpha = 3  Y  \, ,
\ee
\be
r^{\alpha} K^A_{i_A \ov{j}_A \, \alpha} = \lambda_A K^A_{i_A
\ov{j}_A} \, ,
\ee
\be
r^{\alpha} P_{i_A j_B \, \alpha} = {\lambda_A + \lambda_B
\over 2} P_{i_A j_B} \, ,
\ee
\be
r^{\alpha} ({\rm Re \,} f_{ab})_{\alpha} = \lambda_f  {\rm Re
\,}f_{ab} \, ,
\ee
where it is unambiguous to define $Y_\alpha \equiv \partial Y /
(\partial r^{\alpha}) \equiv \partial Y / (\partial z^{\alpha})
\equiv \partial Y / (\partial \ov{z}^{\ov{\alpha}})$, etc.
If there are field configurations such that $w \ne 0$ with
$\cg_i = 0$, there are supersymmetry-breaking minima with
classically vanishing vacuum energy, and eqs.~(\ref{qexpr})
and (\ref{ricci}) give
\be
\label{qfinal}
Q = \sum_A \left( 1+\lambda_A \right) n_A - n - \lambda_f d_f - 1 \,
,
\ee
where $n$ is the number of $z^{\alpha}$ fields, $\sum_A n_A + n =
N_{TOT}$, and $d_f$ is the dimension of the gauge group. From
eq.~(\ref{qfinal}) we can immediately read the contributions to
$Q$ from all multiplets, once their scaling weights are given:
the requirement that $Q=0$, which completes the definition of the
LHC models, amounts to a field-independent but highly non-trivial
constraint.

\begin{center}
{\bf 3. Mass terms in LHC models}
\end{center}

In the case of LHC models, the general supergravity mass formulae and
the resulting expressions for the MSSM mass parameters undergo
dramatic
simplifications [\ref{fkz}] (similar predictions were derived, for
special goldstino directions and under slightly different
assumptions,
in ref.~[\ref{bim}]).

Since the spin-0 fields $z^{\alpha}$ in the supersymmetry-breaking
sector
are assumed here to be gauge singlets with interactions of
gravitational
strength, they have always masses ${\cal O}(m_{3/2}^2/\mpl)$, i.e.
in the $10^{-3}$--$10^{-4} {\rm \; eV}$ range if the gravitino mass
is
at the electroweak scale, with interesting astrophysical
[\ref{astro}] and
cosmological [\ref{cosmol}] implications, including a number of
potential
phenomenological problems. After subtracting the goldstino, their
spin-$1/2$ partners $\chi^{\alpha}$ all have masses equal to the
gravitino
mass.

For the gaugino masses one finds that, if there is unification of the
gauge couplings, $({\rm Re \,} f)_{ab} = \delta_{ab} / g_U^2$, then
\be
\label{mgaugen}
m_{1/2}^2 = \lambda_f^2 \, m_{3/2}^2 \, ,
\;\;\;\;\;
(\lambda_f=0,1) \, .
\ee

As for the spin-$1/2$ fermions $\chi^i$, we should distinguish two
main possibilities. Those in chiral representations of the gauge
group,
such as the quarks and the leptons, cannot have gauge-invariant mass
terms.
Those in real representations of the gauge group, such as the
Higgsino fields of the MSSM, can have both a `superpotential' mass,
proportional to $w_{i_A j_B}$, and a `gravitational' mass,
proportional
to $P_{i_A j_B} [ 1 + (\lambda_A + \lambda_B)/2]$. Both these terms
can
in principle contribute to the superpotential `$\mu$-term' of the
MSSM,
and to the associated off-diagonal (analytic-analytic) scalar mass
term
$m_3^2$. Writing as usual $m_3^2 \equiv B \mu$, one obtains
\be
B = (2 + \lambda_{H_1} + \lambda_{H_2}) m_{3/2}
\ee
in the first case, and
\be
B = \left( 2 + {\lambda_{H_1} + \lambda_{H_2} \over 2} \right)
m_{3/2}
\ee
in the second case.

Moving further to the spin-0 bosons $z^i$ in chiral representations
(squarks, sleptons, \ldots), they can only have diagonal
(analytic-antianalytic) mass terms, of the form
\be
\label{mogen}
(m_0^2)_A = (1+\lambda_A) m_{3/2}^2 \, .
\ee
Similarly, a general formula can be obtained for the
coefficients of the cubic scalar couplings,
\be
(A)_{i_A j_B k_D} = (3 + \lambda_A + \lambda_B + \lambda_D) m_{3/2}
\, .
\ee

The previous discussion should have clarified some important features
of LHC models: the MSSM mass terms are predicted, as functions of the
gravitino mass, by simple formulae involving the approximate scaling
weights; the MSSM $\mu$-term can originate from the K\"ahler
potential,
and is naturally of the order of the gravitino mass; some desirable
universality properties of the MSSM soft mass terms are not
automatic,
but can be ascribed to some universality properties of the
corresponding
scaling weights.

\begin{center}
{\bf 4. Superstring-derived LHC models}
\end{center}

At the pure supergravity level, the assumptions defining the  LHC
models might appear plausible, but they certainly are not compulsory.
It is then remarkable that, if one considers the effective
supergravities
[\ref{efft}] corresponding to the known four-dimensional superstring
models [\ref{fds}], in the appropriate limit one obtains precisely
the
desired scaling properties. The candidate $z^{\alpha}$ fields are the
singlet moduli fields: the universal `dilaton-axion' multiplet $S$,
which entirely determines, at the classical level, the gauge
kinetic function $f_{ab} = \delta_{ab} S$,
and the other moduli that parametrize the size and the shape
of the internal compact space and are usually denoted by the
symbols $T_i$ and $U_i$. In the limit where the $T_i$
and/or $U_i$ moduli are large with respect to the string scale,
the K\"ahler manifold for the chiral superfields displays the
desired properties, with well-defined scaling weights of the
K\"ahler metric with respect to the real combinations $(T_i +
\ov{T}_i)$ and $(U_i+\ov{U}_i)$. These scaling properties are
due to the discrete target-space duality symmetries [\ref{duality}]
of four-dimensional superstrings, and the scaling weights are
nothing but the modular weights with respect to the moduli fields
that participate in the supersymmetry-breaking mechanism: in the
limit of large moduli, non-trivial topological effects on the
world-sheet are exponentially suppressed and can be neglected,
and the discrete duality symmetries are promoted to accidental
scaling symmetries of the kinetic terms in the effective
supergravity theory.

The discrete target-space dualities are
symmetries of the full K\"ahler function $\cg$. Under a generic
duality transformation, of the form $z^\alpha \longrightarrow
f(z^\alpha)$, the K\"ahler potential transforms as $ K
\longrightarrow K + \phi + \ov{\phi}$,
where $\phi$ is an analytic function of the moduli fields $z^\alpha$.
The fact that duality is a symmetry then implies a
definite transformation property for the superpotential, $w
\longrightarrow e^{- \phi} w$. This in turn puts very strong
restrictions on the superpotential modifications that can be
used to describe spontaneous supersymmetry breaking in the
effective supergravity theory. As for the origin of these
superpotential modifications, two types of mechanisms for
supersymmetry breaking have been considered so far in the
framework of four-dimensional string models. The first one
corresponds to exact tree-level string solutions, in which
supersymmetry is broken via orbifold compactifications [\ref{ss}].
The second one is based on the assumption that supersymmetry
breaking is induced by non-perturbative phenomena,
such as gaugino condensation, at the level of the
string effective field theory [\ref{gcond}].

In the string models with tree-level supersymmetry breaking
[\ref{ss}],
the superpotential modifications in the large-moduli limit are fully
under control, since in that case the explicit form of the one-loop
string partition function is known, and one can derive the low-energy
effective theory without making any assumption.
One obtains automatically the desired scaling properties of the
kinetic
terms, which in some cases can produce an LHC model. In this class of
models, the large-moduli limit is a necessity, since for values
of the moduli close to their self-dual points there are
singularities,
induced by some winding modes that become massless and then
tachyonic.
In the large-moduli limit, we can disregard the effects of these
extra
states, excluding them from the effective field theory, and the
superpotential modification associated with supersymmetry breaking
is not manifestly covariant with respect to target-space duality.
On the other hand, the K\"ahler potential maintains the same
expression
as in the case of exact supersymmetry, with the desired scaling
properties
that can produce an LHC supergravity model. Another important
property
of this class of models is the fact that, in order to have $m_{3/2}
\simlt 1 \tev$, some internal radius must be pushed to very large
values.
By dimensional analysis, one would expect that huge threshold
corrections
to the coupling constants, due to the infinite tower of Kaluza-Klein
excitations, spoil the perturbative expansion just above the
compactification scale, but in the framework of string theories this
problem can be avoided.

In the case of non-perturbative supersymmetry breaking [\ref{gcond}],
in the absence
of a second-quantized string formalism one can assume that, at the
level of the effective supergravity, the super-Higgs mechanism is
induced
by a superpotential modification that preserves target-space duality.
Unfortunately, the form of the superpotential modification cannot be
uniquely
fixed by the requirement that it is a modular form of appropriate
weight.
However, another important constraint comes from the physical
requirement
that the potential must break supersymmetry and generate a vacuum
energy at
most ${\cal O}(m_{3/2}^4)$ in the large moduli limit. This is not the
case
for the models of supersymmetry breaking having minima of the
effective
potential for values of the internal moduli all close to some
self-dual
point, and making use of the Dedekind function
$\eta$ in the superpotential modification: either they do not break
supersymmetry or they do so with a large cosmological constant, in
contradiction with the assumption of a constant flat background.
On the other hand, superpotential modifications such that
$w \longrightarrow {\rm constant} \ne 0$ for $z^\alpha \rightarrow
\infty$ give rise to supersymmetry-breaking minima of the effective
potential corresponding to a vacuum energy ${\cal O}(m_{3/2}^4)$ and
$z^{\alpha}$ field configurations far away from the self-dual points.

\begin{center}
{\bf 5. Dynamical determination of $m_Z$ and $m_{3/2}$}
\end{center}

The effective low-energy theories of the LHC supergravity models
contain,
besides the states of the MSSM, the additional light scalars
$z^{\alpha}$
in the supersymmetry-breaking sector, with interactions of
gravitational
strength, whose VEVs control the sliding gravitino mass. This allows
for
a dynamical determination of both $m_Z$ and $m_{3/2}$, via the
logarithmic
quantum corrections associated with renormalizable MSSM interactions.

The conventional treatment of radiative symmetry breaking
[\ref{mssm}]
can be briefly summarized as follows. As a starting point, one
chooses
a set of numerical input values for the independent model parameters
at the unification scale $Q = M_{\rm U}$: the soft masses
($m_0,m_{1/2},A,
m_3^2$), the superpotential mass $\mu$, the unified gauge coupling
$\alpha_U$, and the third-generation\footnote{For the purposes of the
present paper, mixing effects and all other Yukawa couplings can be
neglected.} Yukawa couplings ($\alpha_t,\alpha_b,\alpha_\tau$). One
then  evolves all the running parameters down to a low scale $Q \sim
\msu$, according to the appropriate renormalization group equations
(RGE), and considers the renormalization-group-improved tree-level
potential
\be
\label{vmssm}
V_0(Q) = m_1^2 v_1^2 + m_2^2 v_2^2 + 2 m_3^2 v_1 v_2 + {g'^{\,
2} + g^2 \over 8} \left( v_2^2 - v_1^2 \right)^2 + \dvc \, .
\ee
In eq.~(\ref{vmssm}), $\dvc$ stands for a Higgs-field-independent
contribution to the vacuum energy (cosmological term). The
minimization of the potential in eq.~(\ref{vmssm}), with respect
to the dynamical variables $v_1 \equiv \langle H_1^0 \rangle$ and
$v_2 \equiv \langle H_2^0 \rangle$, is straightforward: for
appropriate numerical assignments of the boundary conditions,
one obtains a phenomenologically acceptable vacuum, with
$SU(2)_L \times U(1)_Y$ broken down to $U(1)_{em}$ and a mass
spectrum compatible with  present experimental data.

Here we regard the MSSM as the low-energy effective theory
of an LHC supergravity model, where the gravitino
mass $m_{3/2}$ cannot be determined at the classical level,
and there are no quantum corrections to the effective potential
carrying positive powers of the cut-off scale $\mpl$. Even if
more general LHC models can be constructed, we assume that some
non-perturbative dynamics fixes the VEVs of the moduli associated
with $\alpha_U$ and $M_{\rm U}$, and the following boundary
conditions on
the MSSM mass parameters
\be
\label{boundary}
m_{1/2} = \xi_1 \cdot m_{3/2} \, ,
\;\;\;\;
m_0 = \xi_2 \cdot m_{3/2} \, ,
\;\;\;\;
A = \xi_3 \cdot m_{3/2} \, ,
\;\;\;\;
m_3 = \xi_4 \cdot m_{3/2} \, ,
\;\;\;\;
\mu = \xi_5 \cdot m_{3/2} \, ,
\ee
where the scaling weights with respect to target-space duality fix
the $\xi$ parameters to constant numerical values ${\cal O} (1)$ or
smaller. Then all the moduli dependence of the MSSM mass parameters
is encoded in the gravitino mass $m_{3/2}$, which should be
considered as an extra dynamical variable, in addition to the Higgs
VEVs $v_1$ and $v_2$. If we take the low-energy limit and neglect the
interactions of gravitational strength, we can formally decouple the
supersymmetry-breaking sector and recover the MSSM. Quantum effects
in
the underlying fundamental theory, however, would induce a
cosmological
term in the resulting MSSM effective potential; for LHC models, this
term contains no positive powers of $\mpl$ and must therefore be
proportional to $m_{3/2}^4$:
\be
\label{cosmo}
\dvc = \eta \cdot m_{3/2}^4 \, ,
\ee
obeying a boundary condition $\eta(M_{\rm U}) = \eta_0$. We stress
that,
in contrast with conventional treatments, in the present context we
are forced to include the cosmological term, since the gravitino mass
is not taken as an external parameter, but rather as a dynamical
variable.

According to our program, one should minimize the effective
potential of the MSSM not only with respect to the Higgs fields,
but also with respect to the new dynamical variable $m_{3/2}$,
keeping (for the moment) the values of $\alpha_U$,
$M_{\rm U}$, $\vec{\xi}$, $\eta_0$ and
$(\atu,\alpha_b^U,\alpha_\tau^U)$
as external input data. As in the standard
approach, the role of radiative corrections is crucial in
developing a non-zero value for the Higgs VEVs at the minimum.
Quantum corrections to the classical potential are summarized, at the
one-loop level, by $V_1 = V_0(Q) + \Delta V_1 (Q)$, where
\be
\label{dv}
\Delta V_1 (Q) = {1 \over 64 \pi^2} {\rm Str} \, \cm^4 \left( \log
{\cm^2 \over Q^2} - {3 \over 2} \right) = \sum_i {n_i m_i^4 \over 64
\pi^2}  \left( \log {m_i^2 \over Q^2} - {3 \over 2} \right) \, ,
\ee
and $V_0(Q)$ is the tree-level potential, eq.~(\ref{vmssm}),
expressed
in terms of renormalized fields and parameters at the scale $Q$. The
RGE for the new dimensionless coupling of the theory, the coefficient
$\eta$ of the cosmological term, reads
\be
\label{etarg}
{d \eta \over d t} = {1 \over 32 \pi^2} \left[ {{\rm Str} \, \cm^4
\over m_{3/2}^4} \right]_{v_{1,2}=0} \, ,
\;\;\;\;\;
(t \equiv \log Q) \, ,
\ee
and plays an important role in the determination of the
supersymmetry-breaking scale: the MSSM particle content is such
that $(\eta-\eta_0)$ is always driven towards negative values at
sufficiently low scales, and the gravitino mass dynamically relaxes
to a value closely related to the scale at which
$\eta$ turns from positive to negative. The desired hierarchy can be
generated for values of $\eta_0$ between zero and ${\cal O}(100)$,
depending on the values of the $\vec{\xi}$ parameters.

To illustrate the main point of our approach, it is convenient to
choose, as independent variables, the supersymmetry-breaking scale
$m_{3/2}^2$ and the dimensionless ratios $\hat{v}_i \equiv
(v_i/m_{3/2})$ ($i=1,2$). Then the minimization condition of the
one-loop effective
potential with respect to $m_{3/2}$ can be written in the form
\be
\label{master}
m_{3/2}^2 {\partial V_1 \over \partial m_{3/2}^2} = 2 V_1 + {{\rm Str
\,} \cm^4 \over 64 \pi^2} = 0 \, .
\ee
The minimization conditions with respect to the variables $\hat{v}_i$
are completely equivalent to the ones that are usually considered in
the MSSM, when the supersymmetry-breaking scale is a fixed numerical
input.
A general study of the MSSM predictions, as functions of the boundary
conditions $\vec\xi$, $\eta_0$, $\alpha^U_{t,b,\tau}$, can be
performed
numerically. Some illustrative numerical results, for a particularly
simple choice of boundary conditions, can be found in
ref.~[\ref{kpz}].

\begin{center}
{\bf 6. Dynamical determination of Yukawa couplings}
\end{center}

We can extend the previous approach by assuming that also
$\alpha^U_{t,b,\tau}$ are dynamical variables [\ref{kpz}]
(see also [\ref{bindud}]), in analogy to what was done before
for the supersymmetry-breaking
scale $\msu \sim m_{3/2}$. The main motivation for this proposal
comes from four-dimensional superstrings, where all the parameters of
the
effective low-energy theory are related to the VEVs of some moduli
fields. If the dynamical mechanism that breaks supersymmetry
fixes the gauge coupling constant $\alpha_U$ to a given numerical
value at $M_{\rm U}$, but leaves a residual moduli dependence of the
Yukawa
couplings, along some approximately flat direction, then also the
latter should be treated as dynamical variables in the low-energy
effective theory. This means that the effective potential of the MSSM
should also be minimized with respect to the moduli on which the
Yukawa
couplings depend.

It is a well-known fact that, in general four-dimensional string
models, tree-level Yukawa couplings are either vanishing or of the
order of the unified gauge coupling. Concentrating for the moment on
the top Yukawa coupling, one expects a tree-level relation of the
form
\be
\label{hyper}
\atu = c_t \alpha_U \, ,
\ee
where $c_t$ is a model-dependent group-theoretical constant of order
unity. At the one-loop level, both gauge and Yukawa couplings receive
in general string threshold corrections [\ref{thr}], induced by the
exchange of Kaluza-Klein and winding states, whose masses depend on
the VEVs of some moduli fields. One can then consider two main
possibilities.

If the top Yukawa coupling receives a string threshold correction
identical to the one of the gauge coupling, the unification
condition (\ref{hyper}) is preserved. In this case the
non-perturbative phenomena, which we have assumed to determine
$\alpha_U$, also fix the value of $\atu$; the latter
is no longer an independent parameter, and one can perform the
analysis described in the previous paragraph with one parameter
less. In particular, the structure of the RGE for $\alpha_t$
is such that its numerical value at the electroweak scale is
always very close to its effective infrared fixed point [\ref{ifp}],
$\alpha_t \simeq 1 /(4 \pi)$.

If eq.~(\ref{hyper}) receives non-trivial threshold corrections, with
additional moduli dependences besides the combination appearing in
the
gravitino mass, it is plausible to assume that also some of the extra
moduli correspond to approximately flat directions, after the
inclusion of
the non-perturbative physics that breaks supersymmetry and fixes the
value
of the unified gauge coupling constant. Then minimization with
respect to
the moduli relevant for the low-energy theory always admits solutions
corresponding to the maximum allowed value for $\alpha_t$. Strictly
speaking, this is excluded by the requirement of perturbative
unification,
$\atu < 1$: the minimum value of the effective potential must
correspond
to the case in which $\alpha_t$ has the largest value permitted by
its
moduli dependence, which is typically very close to the effective
infrared
fixed point.

Both situations described above are very interesting, since they give
the numerical prediction that, at low energy, $m_t \sim M_t^{IR} \sin
\beta$,
where $M_t^{IR} \simeq 190 \gev$ (with an uncertainty of roughly $10
\%$ due
to the error on $\alpha_3$, threshold and higher-loop effects, etc.),
and
$(1/\sqrt{2}) < \sin \beta < 1$ (the actual value being determined by
the
form of the boundary conditions at $M_{\rm U}$), thus in the range at
present allowed by experimental data. In the second case, the reason
of the
attraction of $\alpha_t(m_{3/2})$ towards the infrared fixed point is
the particular structure of the effective potential, after the
minimization
with respect to the supersymmetry-breaking scale. Indeed,
eq.~(\ref{master})
can be rewritten as
\be
\label{explicit}
V_1 |_{min} = - {1 \over 128 \pi^2}
{\rm Str \,} \cm^4  (Q)
= {1 \over 128 \pi^2}
(- m_t^2 C_t^2 - \ldots) \, ,
\ee
where
\be
C_t^2 = 12 \left[ m_{Q_3}^2 + m_{U_3}^2 + {m_Z^2 \over 2} \cos 2
\beta
+ \left( A_t + {\mu \over \tan \beta} \right)^2 \right] \, .
\ee
Equation~(\ref{explicit}) looks unbounded from below in the variable
$\alpha_t$, but the actual bound is set by the effective
infrared fixed point, which therefore corresponds to the deepest
minimum of the effective potential, if permitted by the structure
of the moduli space of the underlying string theory.

Equation~(\ref{explicit}) can be easily generalized [\ref{kprz}]
to the case in which also $\alpha^U_b$ and $\alpha^U_{\tau}$ are
considered as dynamical variables:
\be
\label{explicitbis}
V_1 |_{min} = {1 \over 128 \pi^2}
(- m_t^2 C_t^2 - m_b^2 C_b^2 - m_\tau^2 C_{\tau}^2 + \ldots) \, ,
\ee
where
$C_b^2$ and $C_{\tau}^2$ have expressions similar to that of $C_t^2$.
In this case, the leading dependence of $V_1$ on the Yukawa couplings
$(\alpha_t,\alpha_b,\alpha_{\tau})$ would attract the latter close to
an effective infrared fixed surface; minimization of the vacuum
energy along this surface can be used to determine the expectation
values
of the individual couplings, which can produce acceptable values of
the
third-generation fermion masses in a wide range of the MSSM parameter
space.

As a final remark, we should point out that this dynamical
determination of the low-energy Yukawa couplings is most naturally
embedded in LHC models, but is in principle possible also in models
where both the supersymmetry-breaking scale $m_{3/2}$ and the unified
gauge coupling $\alpha_U$ are fixed by Planck-scale physics, and only
the Yukawa couplings are controlled by the VEVs of singlet scalar
fields
along approximately flat directions of the effective potential.

\begin{center}
{\bf Acknowledgements}
\end{center}
The author would like to thank S.~Ferrara, C.~Kounnas, I.~Pavel and
G.~Ridolfi for instructive and entertaining collaborations.

\newpage

\begin{center}
{\bf References}
\end{center}
\begin{enumerate}
\item
\label{fkz}
S.~Ferrara, C.~Kounnas and F.~Zwirner, preprint CERN-TH.7192/94,
LPTENS-94/12, UCLA/94/TEP13, hep-th/9405188.
\item
\label{kpz}
C.~Kounnas, I.~Pavel and F.~Zwirner, preprint CERN-TH.7185/94,
LPTENS-94/08, hep-ph/9406256.
\item
\label{kprz}
C.~Kounnas, I.~Pavel, G.~Ridolfi and F.~Zwirner, `Possible dynamical
determination of $m_t$, $m_b$ and $m_{\tau}$', CERN preprint, in
preparation.
\item
\label{mssm}
For reviews and references, see, e.g.:
H.-P.~Nilles, Phys. Rep. 110 (1984) 1;
S.~Ferrara, ed., `Supersymmetry' (North-Holland, Amsterdam, 1987);
F.~Zwirner, in `Proceedings of the 1991 Summer School in High Energy
Physics and Cosmology', Trieste, 17 June--9 August, 1991 (E.~Gava,
K.~Narain, S.~Randjbar-Daemi, E.~Sezgin and Q.~Shafi, eds.), Vol.~1,
p.~193.
\item
\label{nscl}
E.~Cremmer, S.~Ferrara, C.~Kounnas and D.V.~Nanopoulos, Phys. Lett.
B133 (1983) 61;
R.~Barbieri, S.~Ferrara and E.~Cremmer, Phys. Lett. B163 (1985) 143.
\item
\label{nsqu}
J.~Ellis, A.B.~Lahanas, D.V.~Nanopoulos and K.~Tamvakis, Phys. Lett.
B134 (1984) 429;
J.~Ellis, C.~Kounnas and D.V.~Nanopoulos, Nucl. Phys. B241 (1984) 406
and B247 (1984) 373.
\item
\label{grk}
M.T.~Grisaru, M.~Rocek and A.~Karlhede, Phys. Lett. B120 (1983) 110.
\item
\label{bim}
A.~Brignole, L.E.~Ib\'a\~nez and C.~Mu\~noz, Madrid preprint
FTUAM-26/93, and references therein.
\item
\label{astro}
T.R.~Taylor and G.~Veneziano, Phys. Lett. B213 (1988) 450;
J.~Ellis, S.~Kalara, K.A.~Olive and C.~Wetterich, Phys. Lett. B228
(1989)
264;
M.~Gasperini, Turin preprint DFTT-03/94;
M.~Gasperini and G.~Veneziano, preprint CERN-TH.7178/94.
\item
\label{cosmol}
G.D.~Coughlan, W.~Fischler, E.W.~Kolb, S.~Raby and G.G.~Ross,
Phys. Lett. B131 (1983) 59;
J.~Ellis, D.V.~Nanopoulos and M.~Quir\'os, Phys. Lett. B174 (1986)
176;
J.~Ellis, N.C.~Tsamis and M.~Voloshin, Phys. Lett. B194 (1987) 291;
R.~Brustein and P.J.~Steinhardt, Phys. Lett. B302 (1993) 196;
B.~de~Carlos, J.A.~Casas, F.~Quevedo and E.~Roulet, Phys. Lett. B318
(1993) 447;
T.~Banks, D.B.~Kaplan and A.E.~Nelson, Phys. Rev. D49 (1994) 779.
\item
\label{efft}
E.~Witten, Phys. Lett. B155 (1985) 151;
S.~Ferrara, C.~Kounnas and M.~Porrati, Phys. Lett. B181 (1986) 263;
S.~Ferrara, L.~Girardello, C.~Kounnas and M.~Porrati, Phys. Lett.
B193 (1987) 368;
I.~Antoniadis, J.~Ellis, E.~Floratos, D.V.~Nanopoulos and T.~Tomaras,
Phys. Lett. B191 (1987) 96;
S.~Ferrara, L.~Girardello, C.~Kounnas and M.~Porrati, Phys. Lett.
B194 (1987) 358;
M.~Cvetic, J.~Louis and B.~Ovrut, Phys. Lett. B206 (1988) 227;
S.~Ferrara and M.~Porrati, Phys. Lett. B216 (1989) 1140;
M.~Cvetic, J.~Molera and B.~Ovrut, Phys. Rev. D40 (1989) 1140;
L.~Dixon, V.~Kaplunovsky and J.~Louis, Nucl. Phys. B329 (1990) 27.
\item
\label{fds}
For a review and references see, e.g.: B.~Schellekens, ed.,
{`Superstring construction'} (North-Holland, Amsterdam, 1989).
\item
\label{duality}
For a review and references see, e.g.:
A.~Giveon, M.~Porrati and E.~Rabinovici, preprint RI-1-94,
NYU-TH-94-01-01, to appear in Physics Reports.
\item
\label{ss}
R.~Rohm, Nucl. Phys. B237 (1984) 553;
C.~Kounnas and M.~Porrati, Nucl. Phys. B310 (1988) 355;
S.~Ferrara, C.~Kounnas, M.~Porrati and F.~Zwirner,
Nucl. Phys. B318 (1989) 75;
M.~Porrati and F.~Zwirner, Nucl. Phys. B326 (1989) 162;
C.~Kounnas and B.~Rostand, Nucl. Phys. B341 (1990) 641;
I.~Antoniadis, Phys. Lett. B246 (1990) 377;
I.~Antoniadis and C.~Kounnas, Phys. Lett. B261 (1991) 369;
I.~Antoniadis, C.~Mu\~noz and M.~Quir\'os, Nucl. Phys. B397 (1993)
515.
\item
\label{gcond}
J.-P.~Derendinger, L.E.~Ib\'a\~nez and H.P.~Nilles, Phys. Lett. B155
(1985) 65;
M.~Dine, R.~Rohm, N.~Seiberg and E.~Witten, Phys. Lett. B156 (1985)
55;
C.~Kounnas and M.~Porrati, Phys. Lett. B191 (1987) 91;
A.~Font, L.E.~Ib\'a\~nez, D.~L\"ust and F.~Quevedo, Phys. Lett.
B245 (1990) 401;
S.~Ferrara, N.~Magnoli, T.R.~Taylor and G.~Veneziano, Phys.
Lett. B245 (1990) 409;
H.-P.~Nilles and M.~Olechowski, Phys. Lett. B248 (1990) 268;
P.~Bin\'etruy and M.K.~Gaillard, Phys. Lett. B253 (1991) 119;
V.~Kaplunovsky and J.~Louis, Austin preprint UTTG-94-1, LMU-TPPW-94-1
(1994);
J.H.~Horne and G.~Moore, Yale preprint YCTP-P2-94.
\item
\label{bindud}
Y.~Nambu, as quoted by T.~Gherghetta at the IFT Workshop on Yukawa
Couplings, Gainesville, 11--13 February 1994, to appear in the
Proceedings; P.~Bin\'etruy and E.~Dudas, Orsay preprint LPTHE 94/35.
\item
\label{thr}
V.S.~Kaplunovsky, Nucl. Phys. B307 (1988) 145;
L.J.~Dixon, V.S.~Kaplunovsky and J.~Louis, Nucl. Phys. B355 (1991)
649;
J.-P.~Derendinger, S.~Ferrara, C.~Kounnas and F.~Zwirner, Nucl. Phys.
B372 (1992) 145 and Phys. Lett. B271 (1991) 307;
G.~Lopez Cardoso and B.A.~Ovrut, Nucl. Phys. B369 (1992) 351;
I.~Antoniadis, K.S.~Narain and T.~Taylor, Phys. Lett. B276 (1991) 37;
I.~Antoniadis, E.~Gava, K.S.~Narain and T.R.~Taylor, Nucl. Phys. B407
(1993) 706.
\item
\label{ifp}
B. Pendleton and G.G. Ross, Phys. Lett. B98 (1981) 291;
C. Hill, Phys. Rev. D24 (1981) 691;
L.~Alvarez-Gaum\'e, J.~Polchinski and M.B.~Wise, Nucl. Phys. B221
(1983) 495;
J. Bagger, S. Dimopoulos and E. Mass\'o, Phys. Rev. Lett. 55 (1985)
920.
\end{enumerate}
\end{document}